\documentclass[aps,nofootinbib]{revtex4}

\usepackage{amsmath,amssymb,graphicx,float}
\usepackage{color}
\usepackage{textcomp}
\usepackage[normalem]{ulem}

\newcommand{\mathsym}[1]{{}}
\newcommand{\unicode}[1]{{}}

\begin{document}

\title{Exploring the landscape of community-based dismantling strategies}
\author{F. Musciotto, S. Miccich\`e}
\affiliation{Dipartimento di Fisica e Chimica - Emilio Segr\`e, Universit\`a degli Studi di Palermo, Viale delle Scienze, Ed. 18, 90128, Palermo, Italy}

\date{\today}

\begin{abstract}

Network dismantling is a relevant research area in network science, gathering attention both from a theoretical and an operational point of view. Here, we propose a general framework for dismantling that prioritizes the removal of nodes that bridge together different network communities.
The strategies we detect are not unique, as they depend on the specific realization of the community detection algorithm considered. However, when applying the methodology to some real-world networks we find that the percolation threshold at which dismantling occurs is strongly robust, and it does not depend on the specific algorithm. Thus, the stochasticity inherently present in many community detection algorithms allows to identify several strategies that have comparable effectiveness but require the removal of distinct subsets of nodes. This feature is highly relevant in operational contexts in which the removal of nodes is costly and allows to identify the least expensive strategy that still holds high effectiveness. 



\end{abstract}

\maketitle

\section{Introduction}

Network dismantling has become a topic of great interest in many research and operational fields. The basics of dismantling are grounded in the seminal works of Cohen \cite{havlin} and Albert \cite{barabasi} that were the first ones to study the resilience properties of networks. These two works have greatly elucidated the resilience mechanisms of networks under random \cite{havlin} and targeted \cite{barabasi} attacks. Also, they indicated how percolation-based methodologies can be fruitfully used to investigate networks properties. 

Indeed, network dismantling can be seen as the reverse of network resilience, with the {\em trait d'union} being the percolation phenomenon. Indeed, reversing the mechanism of percolation we observe how large connected components of a network abruptly disappear through the progressive deletion of links or nodes, ranked according to some criterion \cite{strog}. Different ranking criteria have been proposed to maximize the efficiency of network dismantling, focusing on the detection of the most influential nodes to remove \cite{venti, venti1, venti2, venti3}, or adopting holistic approaches that focus on the collective, emergent features of complex networks \cite{venti4, venti5, venti6, venti7}. Due to its flexibility, network dismantling has been applied to different domains, ranging from biology \cite{venti8} to socio-technical systems \cite{venti9,bottlenecks} and crime \cite{MafiaDismantling, fiumarademeo}. 

In Ref. \cite{MafiaDismantling} we have considered a dismantling methodology that is based on the membership of mafia affiliates to specific Mafia syndicates. Specifically, we show how prioritizing the removal of nodes that have the highest number of connections with members coming from different syndicates guarantees good dismantling performances. The aim of the present work is to generalize that approach by exploring the idea of dismantling a network with a strategy modeled on its community structure. Specifically, we prioritize the removal of nodes that have the highest number of links with nodes in other communities. Indeed, these nodes are more likely to bridge different areas of the network. We observe how the community can be given (as in our previous investigation of a Mafia network), or can be detected using one of the several methods available in literature~\cite{fortunato2016community}. In the latter case, there is an unavoidable degree of arbitrariness in the choice of the community detection algorithm as well as to the stochastic nature inherited in most of such algorithms. 

This is an issue that has been extensively investigated in a recent paper \cite{zanin}. To our knowledge this is the first paper where community-based dismantling approaches have been dealt with.  In Ref. \cite{zanin} it is shown that community-based network dismantling significantly outperforms existing techniques in terms of solution quality and computation time in the vast majority of the analysed real-world networks, while standard dismantling techniques mainly excel on model networks. The dismantling strategy we propose is different from the one proposed in Ref. \cite{zanin} mainly due to the fact that we directly select nodes to be removed, rather than links. We therefore consider a node percolation dismantling approach rather than a link-percolation approach.

Hereafter, we show that our methodology outperforms methods based on a node removal determined by the computation of nodes with highest degree. However, we find that the community based dismantling strategy is not unique, being it dependant on the specific network partitioning considered. However, rather than being a flaw, we believe that this is a strength of our methodology. In fact, since we are interested in dismantling -- and not in resilience -- having different dismantling paths can be an advantage in all operational situations in which a specific set of nodes to be removed might be unreachable for several reasons, mainly related to their accessibility and/or the removal costs. However, when providing different strategies, we are interested in the stability of the percolation threshold at which dismantling occurs, which is related to the number of nodes that need to be removed from the network before considering it dismantled. In fact, we show that at least for the cases accounted for in this work, such threshold remains essentially the same on all possible strategies, notwithstanding the community detection algorithm. Moreover, we find that the effectiveness of our strategy depends on the significance of the community structure: in random networks, our strategy performs only slightly better than a degree based one.

The paper is structured as follows: In section \ref{results} we apply our approach to three different sets of real world networks and a class of randomly generated ones and in section \ref{concl} we discuss the main implications. In section \ref{model} we briefly sketch the methodology. 

\section{Results} \label{results}

\subsection{The actors-movies dataset} \label{data1}

We apply our method to the set of male actors playing in movies indexed in the IMDb database (http://www.imdb.com/). IMDb is the largest web repository of world movies. We consider here the bipartite relationship between movies and actors produced in the period 1990–2009 all over the world. The set includes movies realized in 169 countries . 

This is a large bipartite network with 89,605 movies and 412,143 male and female actors. From this information we generate the univariate projected network of actors, by setting a link between any two pair of actors whether they played in the same movie \cite{SVN}.

\begin{figure} [H]
\begin{center}
                    \includegraphics[scale=0.55]{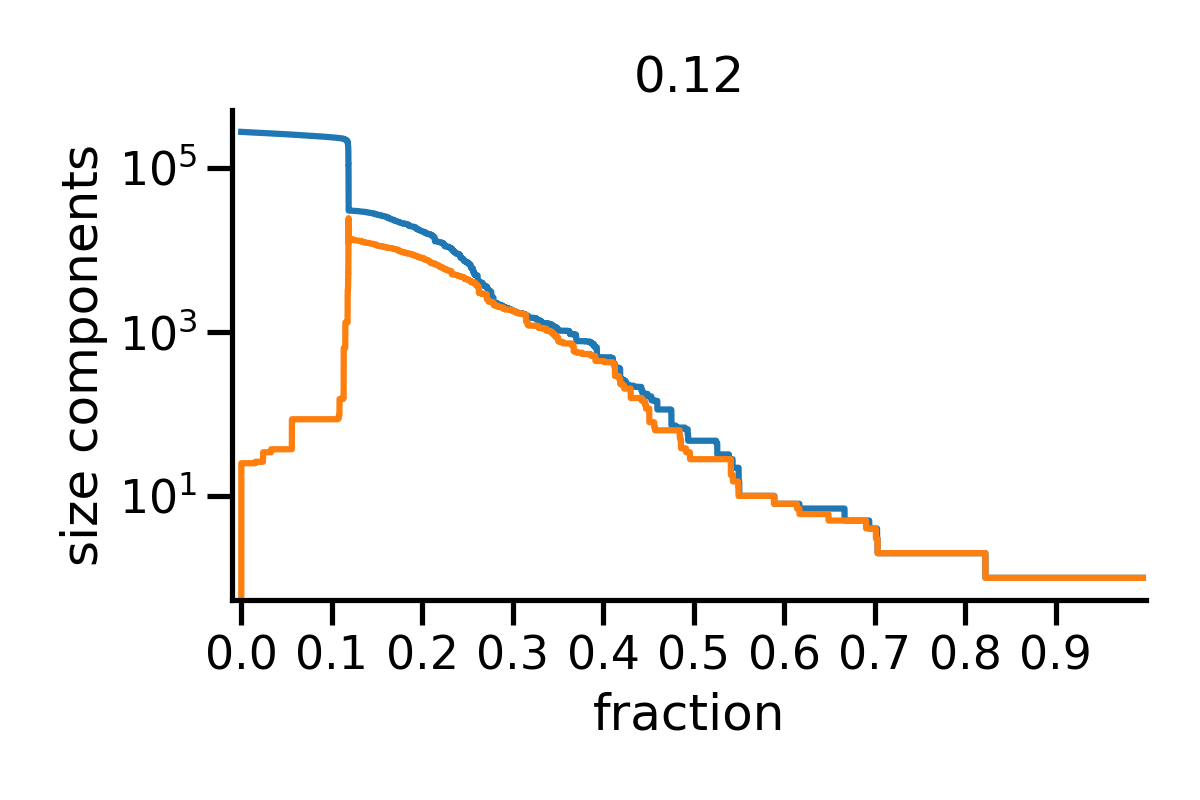}
                    \includegraphics[scale=0.55]{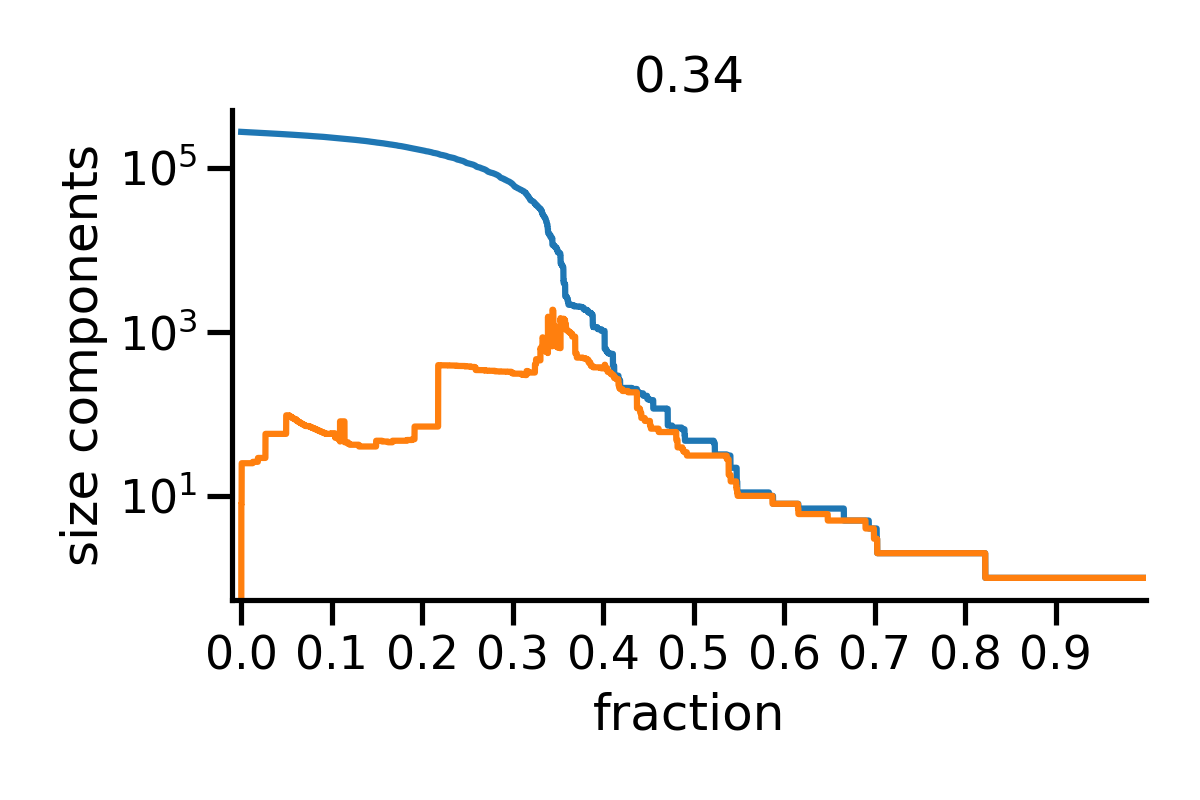}
                    \caption{Sizes of the largest (LCC) and second largest (SLC) connected components as a function of the fraction of removed nodes for the complete IMDB network, using the community based strategy - Leiden algorithm (left panel) and the degree (right panel)} \label{actornet}
\end{center}
\end{figure}

In what follows, we limit our analysis to male actors, thus considering a network made of 274,507 nodes and 6,408,592 links. In the left panel of Fig. \ref{actornet} we show the size of the largest connected component (LCC) and second largest connected component (SLC) as a function of the fraction of nodes removed according to the procedure illustrated in section \ref{model}. When the fraction increases, the difference between the size of the LCC and the size of the SLC gets smaller. This is a clear indication of the fact that while originally all nodes are connected in a single giant component, as soon as the removal process takes place, a number of smaller connected components start emerging and the network gets fragmented. When increasing the number of removed nodes we see the formation of smaller components of increasing size at the expense of the largest component. However, as soon as the percentage $q$ of removed nodes exceeds a certain critical value we observe that all these smaller components become much more numerous and much smaller. Thus, we need to define a critical threshold for the fraction of removed nodes $q_c$, after which we consider the network to be dismantled. Taking inspiration from existing literature~\cite{barabasi,MafiaDismantling}, we set this threshold as the value of $q$ that maximizes the size of the SLC. The critical threshold $q_c^{(comm)}$ after which the networks gets dismantled is $q_c^{(comm)} \approx 0.12$. The right panel of Fig. \ref{actornet} show the same analysis for the case when the removal of nodes is done according to their degree. Although the behaviour is qualitatively the same, one can notice that the critical threshold $q_c^{(deg)}$ after which the networks gets dismantled is $q_c^{{deg}} \approx 0.34$, which is significantly higher than $q_c^{{comm}}$.

In order to test how robust is our procedure, we repeat the whole procedure M times. However, due to the large computational effort needed for the dismantling of the whole network (11 days on an Intel$^\circledR$  based workstation
Intel$^\circledR$ Xeon$^\circledR$ Gold 5118 CPU $@$ 2.30GHz - 2 processors -  total 24 cores 512 Gb RAM) we decided to split the network according to the production year of the movies. We then split the network into 20 smaller networks, one per year in the interval 1990-2009. In Table \ref{tabactornet} we show, for each yearly network, the number of nodes (second column), the number of links (third column), the mean and standard deviations of the $q_c^{(comm)}$ thresholds (fourth column) computed by repeating our procedure M=30 times and the $q_c^{(deg)}$ thresholds (fifth column) computed by considering the usual degree-based node removal. We observe that our methodology is quite robust, as the dispersion of the $q_c^{(comm)}$ threshold values around their mean is below $1 \%$ of the mean. Moreover, in all 20 considered networks, $q_c^{(comm)}$ is smaller than $q_c^{(deg)}$. This clearly show how our methodology, at least for this network, is robust and even outperforming standard methodologies based on the removal of nodes with the highest degree.
\begin{table} [H]
\begin{center}
\begin{tabular}{| c | c | c | c | c |c |}
\hline
          year  & nodes  ~&~links   ~&~$q_c^{(comm)}$     ~&~$q_c^{(deg)}$~\cr \hline
          2009 & 34784    & 431074 & 0.0871 $\pm$ 0.0007 & 0.23                  \cr
          2008 & 35287    & 463305 & 0.0618 $\pm$ 0.0007 & 0.21                  \cr
          2007 & 32948    & 455193 & 0.0628 $\pm$ 0.0006 & 0.19                  \cr
          2006 & 33930    & 472873 & 0.0578 $\pm$ 0.0005 & 0.18                  \cr
          2005 & 30936    &  443281            & 0.0568 $\pm$ 0.0010 & 0.18                 \cr
          2004 & 27532    &  362056            & 0.0467 $\pm$ 0.0005 & 0.16                 \cr
          2003 & 23874    &  352889            & 0.0463 $\pm$ 0.0007 & 0.17                 \cr
          2002 & 21190    &  309913            & 0.0611 $\pm$ 0.0012 & 0.18                 \cr
          2001 & 21708    &  313105            & 0.0605 $\pm$ 0.0009 & 0.20                 \cr
          2000 & 19894    &  292269            & 0.0602 $\pm$ 0.0007 & 0.18                 \cr
          1999 & 19915    &  296424            & 0.0586 $\pm$ 0.0008 & 0.10                 \cr
          1998 & 17848    &  258265            & 0.0670 $\pm$ 0.0005 & 0.18                 \cr
          1997 & 17256    &  260975            & 0.0717 $\pm$ 0.0010 & 0.20                 \cr
          1996 & 16451    &  259303            & 0.0649 $\pm$ 0.0008 & 0.19                 \cr
          1995 & 14918    &  219909            & 0.0638 $\pm$ 0.0010 & 0.19                 \cr
          1994 & 14373    &  225495            & 0.0646 $\pm$ 0.0011 & 0.13                 \cr
          1993 & 13855    &  209918            & 0.0565 $\pm$ 0.0020 & 0.13                 \cr
          1992 & 13907    &  218349            & 0.0513 $\pm$ 0.0006 & 0.18                 \cr
          1991 & 13626    &  203889            & 0.0522 $\pm$ 0.0007 & 0.17                 \cr
          1990 & 13402    &  191827            & 0.0565 $\pm$ 0.0012 & 0.17                 \cr \hline
\end{tabular}
\caption{Summary statistics of the IMDB networks of actors year by year. The columns report (i) the release year of the movies on which the network has been constructed, (ii) the number of nodes, (iii) the number of links, (iv) the critical percentage of removed nodes $q_c^{(comm)}$ using the community based strategy and (v) the critical percentage of removed nodes $q_c^{(deg)}$ using the degree based strategy.} \label{tabactornet}
\end{center}. 
\end{table} 

\begin{figure} [H]
\begin{center}
                    \includegraphics[scale=0.61]{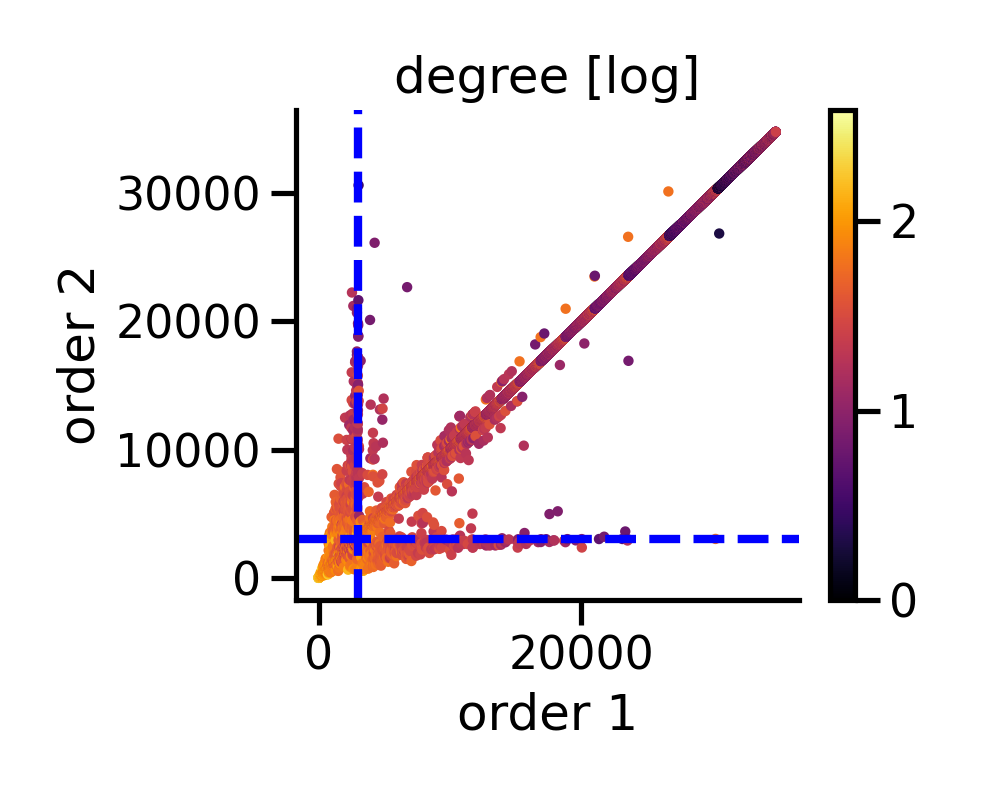}
                    \includegraphics[scale=0.55]{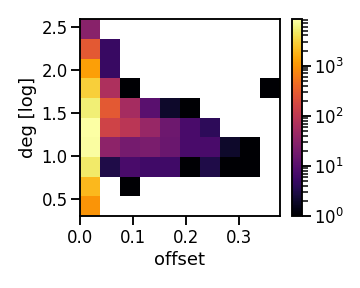}
                    \caption{Left panel: scatter plots of the rankings of nodes in two different realizations of the dismantling procedure based on the community structure of the 2009 IMDB network. The color of nodes reflect their degree (logarithmic scale) while the two dashed blue lines represent the number of nodes that correspond to critical threshold $q_c^{(comm)}$ of network dismantling. Right panel: 2D histogram of the average offset between rankings across all couples of the $M$ realizations of the dismantling procedure (x-axis) versus logarithm of the degree (y-axis). Colors represent the number of counts in each bin.} \label{actornetrob}
\end{center}
\end{figure}

In spite of the robustness of $q_c^{(comm)}$ across several realizations, we highlight that each realization gives different sequences of node removals. Thus, our approach is able to identify several dismantling strategies that have the same effectiveness but involve different nodes. In order to highlight the difference among strategies, in the left panel of Fig. \ref{actornetrob} we show a scatterplot reporting the node ranks obtained in two different realizations of our methodology to the 2009 network. Each node is coloured according to its log degree in the original network - the color code is shown in the palette at the right of the panel. This figure shows that nodes with high (original) degree are removed always in (almost) the same order, as they are mostly located on the diagonal line. Off-diagonal dots, that represents nodes that are removed in significantly different order in the two realizations of the dismantling procedure, have usually darker colors, thus indicating that they have low degree. The blue lines in the panel indicate the number of nodes to be removed in order to dismantle the network, which correspond to the two $q_c^{(comm)}$. All the points that are in the two rectangles up and right the small square attached to the origin represent nodes that are removed in one of the strategies but not in the other one, thus giving rise to different dismantling strategies. The left panel of Figure~\ref{actornetrob} illustrates the comparison between only two of the $M=30$ realizations of the dismantling procedure. In order to obtain a more comprehensive view we consider, for each node in the network, the absolute value of the relative difference between the ranking positions in all couples of the $M$ different applications of our procedure, which we call offset. Specifically, given two dismantling rankings $r^1$ and $r^2$, the offset $o$ for a generic node $i$ is defined as $o_i=\frac{|r^1_i-r^2_i|}{N}$, where $N$ is the number of nodes in the network. In the right panel we report on the horizontal axis such binned offsets and on the vertical axis the binned log degree. At each bin is therefore assigned a number between 0 and $M(M-1)/2$. Such numbers are shown according to the color code reported in the palette at the right of the panel. One can easily see that larger offsets occur less frequently (darker colors) and for nodes with degree of a few units, i.e. between $e^{0.8} \approx 2$ and $e^{1.8} \approx 6$.

The above results show that the intrinsic stochasticity associated to any community detection algorithms clearly affects the sequence of the nodes that have to be removed in order to dismantle the considered network. However, the procedure remains robust with respect the critical threshold $q_c^{(comm)}$. This would suggest that indeed $q_c^{(comm)}$ refers to a peculiar property of the network rather than of the way in which we partition it. Indeed,  when considering the Louvain community detection methodology \cite{louvain} we get a value of  $0.0901 \pm 0.0007$ for the 2009 actors network, which is very close the one obtained with the Leiden clustering.

\subsection{The airport network} \label{data2}

In order to test this last hypothesis we consider in this section a smaller dataset that will be partitioned with different community detection algorithms all based on the maximization of modularity.

Specifically we will consider here a network of $N=1390$ nodes and $L=9758$ symmetric links distributed along 14 connected components whole largest size is $1363$. Nodes are airports of the ECAC area and links between airports are established whenever there is a flight that connects them \cite{elsa, bong, diecid, motifs}.  Our dataset comprises all the flights that, even partly, cross the ECAC airspace for the entire 2017 year. Data were obtained by EUROCONTROL (http://www.eurocontrol.int), the European public institution that coordinates and plans air traffic control for all of Europe. Specifically, we obtained access to the Demand Data Repository (DDR) \cite{ddr2} from which one can obtain all flights followed by any aircraft in the ECAC airspace. Data about flights contain several types of information. In the present study, we just focus on the origin–destination of each flight crossing the ECAC airspace. The specific network we will consider here refers to fligths occurring in day $1_{st}$ September 2017.

\begin{figure} [H]
\begin{center}
                    \includegraphics[scale=0.55]{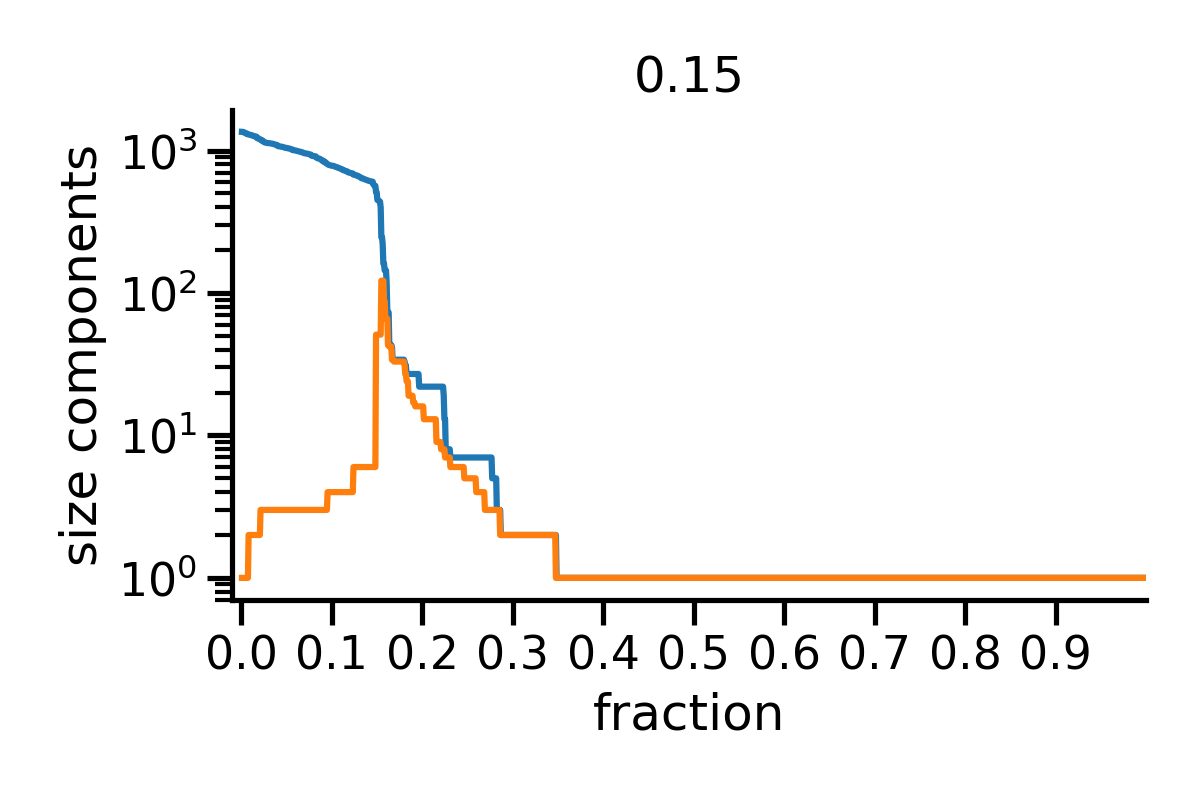}
                    \includegraphics[scale=0.55]{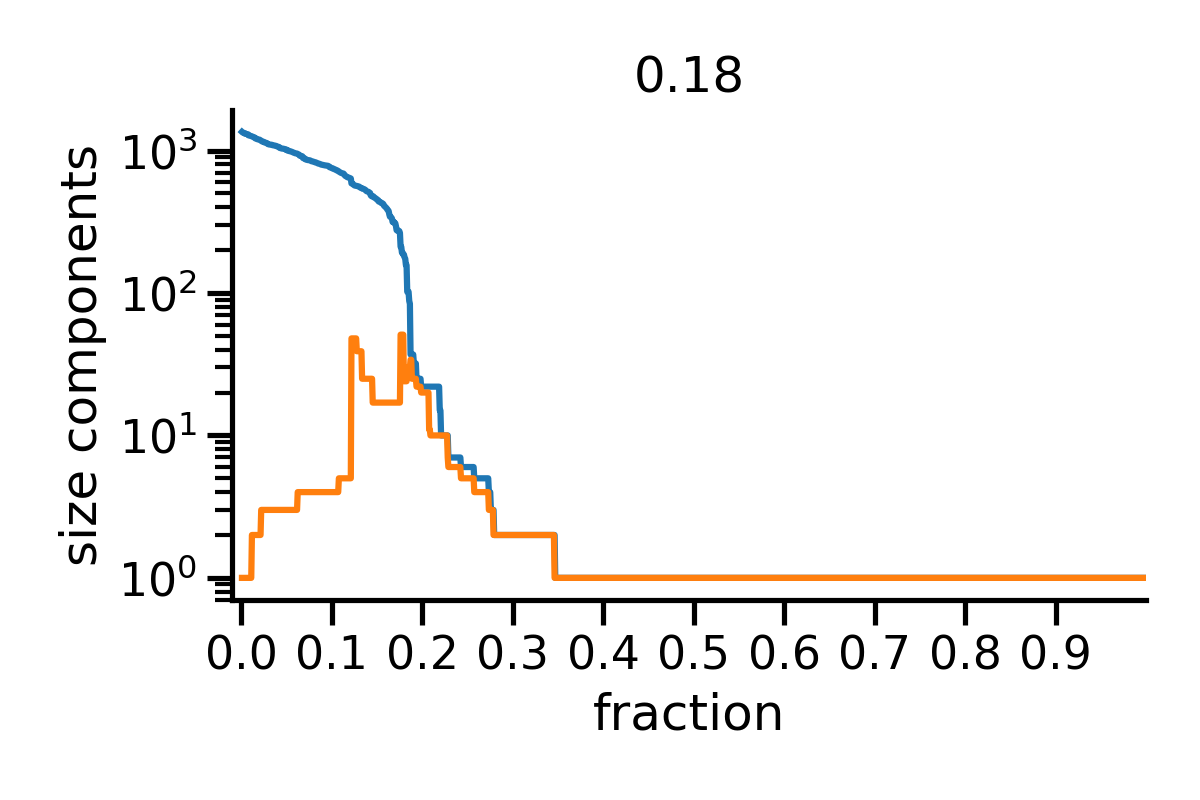}
                    \caption{Sizes of the largest (LCC) and second largest (SLC) connected components as a function of the fraction of removed nodes for the flights network, using the community based strategy - Leiden algorithm (left panel) and the degree (right panel)} \label{airpnet}
\end{center}
\end{figure}

In the left panel of Fig. \ref{airpnet} we show the LCC and the SLC as long as the nodes are removed according to the procedure illustrated in section \ref{model}. The critical threshold $q_c^{(comm)}$ after which the networks gets dismantled is $q_c^{(comm)} \approx 0.15$. The right panel of Fig. \ref{airpnet} show the same analysis for the case when the removal of nodes is done according to their degree. The critical threshold $q_c^{(deg)}$ after which the networks gets dismantled is $q_c^{(deg)} \approx 0.18$, slightly higher than $q_c^{(comm)}$. When applying the procedure $M=100$ times we get and average value of $\langle q_c^{(comm)} \rangle =0.153$ with a standard deviation $\sqrt{\langle [q_c^{(comm)} -\langle q_c^{(comm)} \rangle ]^2 \rangle} =0.007$. It is worth mentioning that when considering the Louvain community detection methodology \cite{louvain} we get an average value of $\langle {\bf{q}}_c^{(comm)} \rangle =0.145$ with a standard deviation $\sqrt{\langle [{\bf{q}}_c^{(comm)} -\langle {\bf{q}}_c^{(comm)} \rangle ]^2 \rangle} =0.006$. The two methodologies give very close results, thus confirming the idea that the value of $q_c^{(comm)}$ seems to refer to a peculiar property of the network rather than of the way in which we partition it.


\subsection{The Email network} \label{data3}

In what follows we replicate the analysis of section \ref{data2} on a different dataset, usually used as a reference benchmark for community detection as it comes with ground truth community structure~\cite{yin2017local, email}.

Specifically we will consider here a network of $N=1005$ nodes and $L=16687$ symmetric links distributed along 20 connected components whole largest size is $986$. Nodes are researcher in a European institution and links are email sent between them. The given community structure reflect the partition in research departments.

\begin{figure} [H]
\begin{center}
                    \includegraphics[scale=0.55]{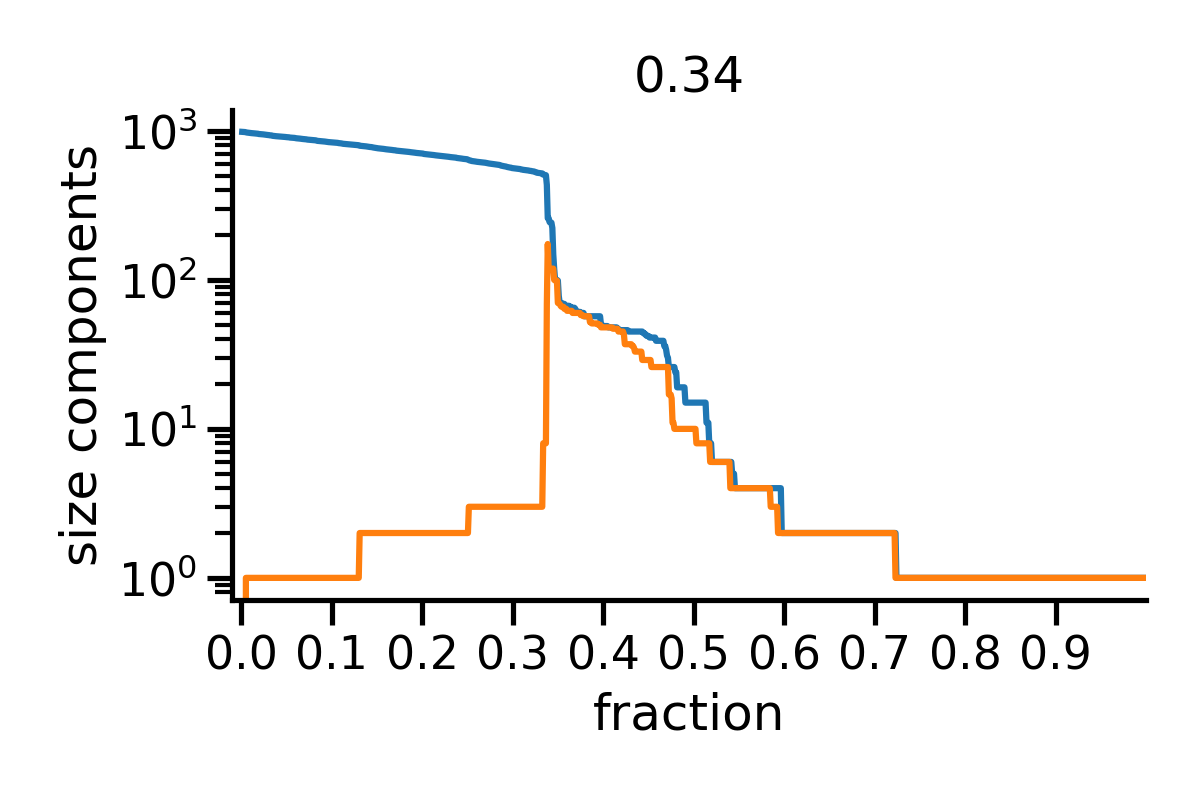}
                    \includegraphics[scale=0.55]{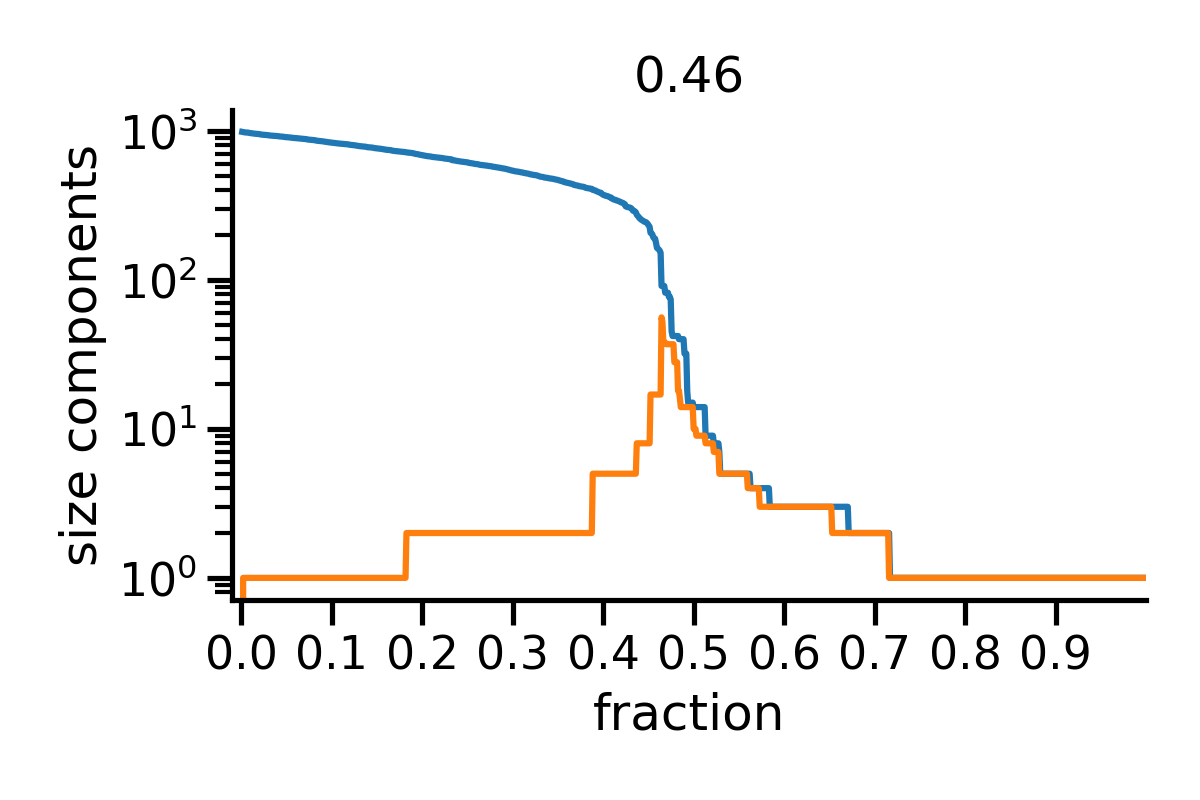}
                    \caption{Size of the largest (LCC) and second largest (SLC) connected components as a function of the fraction of removed nodes for the email network, using the community based strategy - Leiden algorithm (left panel) and the degree (right panel)} \label{emailnet}
\end{center}
\end{figure}

In the left panel of Fig. \ref{airpnet} we show the LCC and the SLC as long as the nodes are removed according to the procedure illustrated in section \ref{model}. The critical threshold $q_c^{(comm)}$ after which the networks gets dismantled is $q_c^{(comm)} \approx 0.34$. The right panel of Fig. \ref{emailnet} show the same analysis for the case when the removal of nodes is done according to their degree. The critical threshold $q_c^{(deg)}$ after which the networks gets dismantled is $q_c^{(deg)} \approx 0.46$, slightly higher than $q_c^{(comm)}$. When applying the procedure $M=100$ times we get an average value of $\langle q_c^{(comm)} \rangle =0.340$ with a standard deviation $\sqrt{\langle [q_c^{(comm)} -\langle q_c^{(comm)} \rangle ]^2 \rangle} =0.005$. Also in this case, when considering the Louvain community detection methodology \cite{louvain} we get an average value of $\langle {\bf{q}}_c^{(comm)} \rangle =0.341$ with a standard deviation $\sqrt{\langle [{\bf{q}}_c^{(comm)} -\langle {\bf{q}}_c^{(comm)} \rangle ]^2 \rangle} =0.004$. Again, the two methodologies give very close results.

The systems we are considering in this section has size comparable with that of the systems analyzed in section \ref{data2}. However, the links are nearly double, which makes the communities of the present system much more interconnected with each other than in the previous case. This is well shown by the fact that the degree-based removal requires that almost 50 \% of the nodes are removed before dismantling. In this case our procedure is clearly better, as the fraction of nodes to be removed gets down to 34\%. By the way this is a percentage similar to the one obtainable by considering a betweenness-based node removal.

Let us now consider a further partitioning of this system. In fact, we consider here an endogenous partitioning given by the membership of the considered individuale (nodes) to one of the 42 departments. In this case we consider departments as communities and nodes are partitioned into 42 communities of sizes ranging from 2 to 109. When considering the Louvain methodology, the average largest size is $226 \pm 23$,  the average smallest size is $56 \pm 7$ with an average number of communities which is about 8. When considering the Leiden methodology, the average largest size is $182 \pm 4$,  the average smallest size is $32.7 \pm 0.5$ with an average number of communities which is again about 8. It is therefore clear that the partitioning in departments is much finer than those obtained with topology based community detection algorithms. However, the dismantling procedure based on the community structure given by the firm's departments gives a percolation threshold of 0.40. This might suggest that the information flow captured by the email exchange does not reflect the departmental structure of the firm. Rather, such flow is better captured by community detection algorithms based on the maximization of modularity.
\begin{figure} [H]
\begin{center}
                    \includegraphics[scale=0.55]{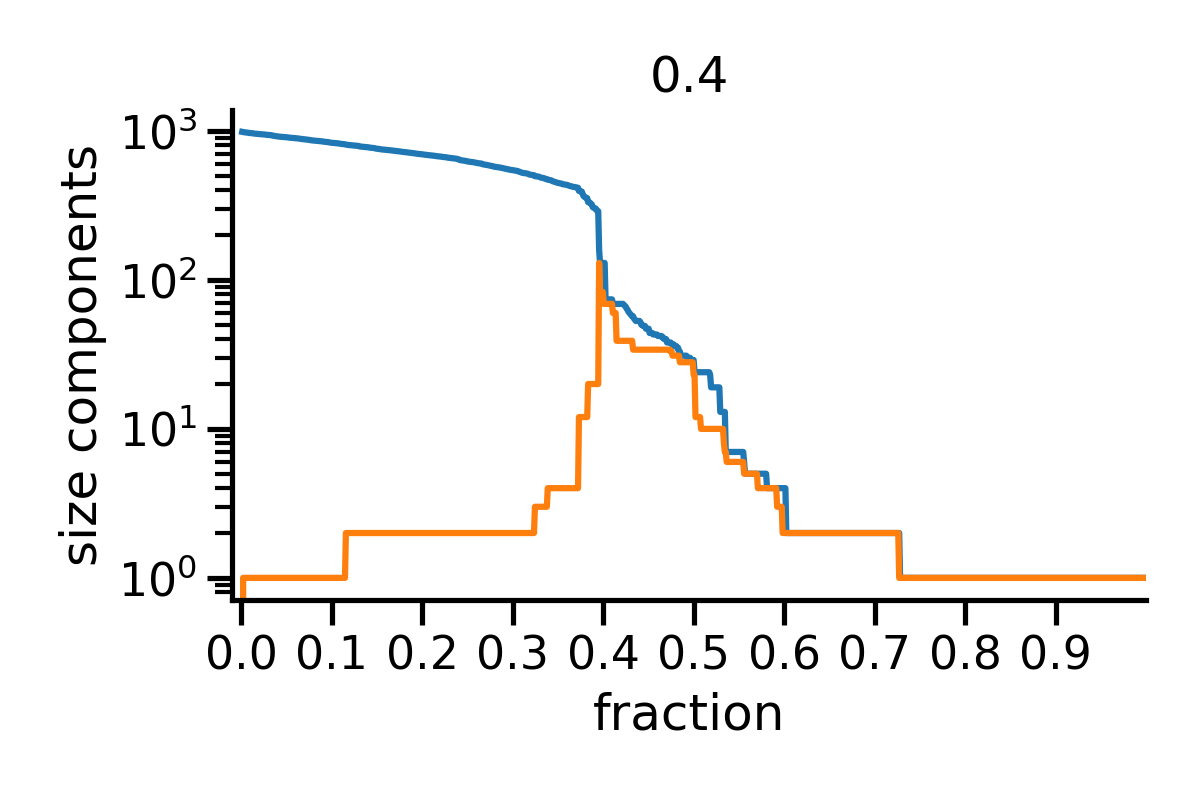}
                    \caption{Size of the largest (LCC) and second largest (SLC) connected components as a function of the fraction of removed nodes for the complete IMDB network, using a dismantling procedure based on the existing community structure.} \label{emailnetDA}
\end{center}
\end{figure}

\subsection{Random networks} \label{data4}

In order to understand if and how the effectiveness of our community based dismantling strategy depends on the properties of the considered networks, we generated a class of random networks using the Barabasi-Albert (BA) generative model. In fact, through this model we are able to reproduce an ubiquitous property of real world networks, which is the presence of a scale-free degree distribution. At the same time, though, the generative process of the BA model is random and does not implant any significant community structure among nodes. Thus, by creating a series of random networks with scale-free degree distributions, we are able to test the performance of our dismantling procedure in the absence of a genuine community structure. To do so, we create random networks with $n=500$ nodes but we vary the number of links $l$ that are added in the BA model when a new node is added, $l\in[2,4,6,8,10]$. For each generated network, we run (i) $M=100$ times our dismantling procedure and (ii) the degree based one, and check the results, see Table~\ref{tab:random}. We observe that when the density increase, as expected the critical threshold also increases in both methods. We still observe that, for all values of $l$, the community based strategy performs slightly better than the degree based one. However, we also find that the relative difference in $q_c$ between the two approaches is reduced if compared to previous real world networks, as the ratio between  $q_c^{(comm)}$ and $q_c^{(deg)}$ fluctuates around 0.9 among the different realizations of random networks. Thus, when the network is lacking a strong, significant community structure, our method is not significantly better than a degree based approach. Indeed, the intuition behind our proposal is that nodes that maximize the connections with other communities play a relevant role in making the system connected. When the community structure is comparable with random noise, the nodes we identify do not have significantly higher dismantling power than those with a high degree.

\begin{table} [H]
\begin{center}
\begin{tabular}{| c | c | c | c | c |}
\hline
          l  & density  ~&n communities & ~$q_c^{(comm)}$     ~&~$q_c^{(deg)}$~\cr \hline
          2  & 0.008   & 14.00 & 0.134 $\pm$ 0.006    & 0.154  \cr
          4  & 0.015   & 11.09 & 0.292 $\pm$ 0.006 & 0.32 \cr 
          6  & 0.023   & 10.49 & 0.402 $\pm$ 0.006 & 0.404  \cr
          8  & 0.031   & 10.00 & 0.472 $\pm$ 0.005 & 0.484 \cr 
          10 & 0.039   & 10.00 & 0.527 $\pm$ 0.006 & 0.558  \cr
          
          \hline
\end{tabular}
\caption{Results of dismantling on randomly generated networks. The columns report (i) $l$ parameter of the Barabasi-Albert generated network, (ii) density of the network, (iii) average number of communities detected by the Leiden algorithm over 100 iterations, (iv) $q_c^{(comm)}$ using the community based strategy and (v) the critical percentage of removed nodes $q_c^{(deg)}$ using the degree based approach} \label{tab:random}
\end{center}. 
\end{table} 

\section{Discussion} \label{concl}

Network dismantling has a twofold importance in the research stream on complex networks. On one side, it is related to the theoretical investigation of the resilience and the percolation properties of complex networks, and as such it helps in shredding light on the structural properties of several real world systems~\cite{havlin,barabasi}. On the other, it holds an operational relevance in more applied contexts of investigation and anti-terrorism agencies~\cite{MafiaDismantling}. In this paper, we have expanded on existing work~\cite{zanin} to show the effectiveness of a dismantling procedure based on the community structure of a network. In fact, as communities represent distinct subsets of the nodes of a network which are characterized by higher inner connectivity, prioritizing the removal of nodes that bridges different communities is an effective way of rapidly dismantling the network. Indeed, we show how such a strategy proves its effectiveness in several different scenarios, always overperforming traditional dismantling strategies based on the centrality of nodes. Moreover, we show how the procedure that we designed has an unavoidable degree of stochasticity, as it is based on non-deterministic algorithm of community detection. With respect to this, we find that in spite of the differences in the outcomes of distinct realizations of the dismantling procedure, the effectiveness of each realization is strongly robust and does not vary significantly. Thus, our approach is able to identify several strategies that require the removal of distinct sets of nodes but all share a similar effectiveness. From an operational point of view, this result has strong implication: it allows to change the nodes that need to be removed to actually dismantle a network, allowing to find sets of nodes that minimize the removal costs without deteriorating the global effectiveness of the strategy. Moreover, the robustness of the results also across different community detection algorithms suggest that the critical threshold that quantify the percentage of nodes that need to be removed to dismantle the network might be a network property, rather than being related to the specific dismantling path. Indeed, we find that the effectiveness of the strategy (and thus the value of the critical thresholds) depends on the significance of the community structure of the network, being lower when the communities are not statistically significant.

\section{Methods} \label{model}

Resilience properties of networks have been firstly studied within the context of random attacks\cite{havlin}, showing that real networks, due to their specific properties (the presence of heterogeneous degree distributions among others) are strongly robust against dismantling strategies where the order of removal is random. Conversely, it has later been shown that targeted attacks, in which nodes are removed according their centrality, are much faster in dismantling a network\cite{barabasi}. In both cases percolation techniques provide the theoretical background  and the operative protocols for understanding how node removal affects network resilience.

Our methodology generalizes the framework of network dismantling by considering a set of removal strategies of nodes that are based on their membership in non-overlapping communities \cite{zanin}. Specifically, nodes will be removed from the network starting from the nodes that have the highest number of links with nodes  of other communities. 

We therefore propose an iterative procedure that at each iteration involves the following four steps:
\begin{itemize}
\item Partition the network according to your preferred community detection algorithm
\item Select the largest community
\item Select the node with the highest number of links with nodes of other communities
\item Remove the selected node
\end{itemize}
Starting from the original network, the iterations proceed until we are left with a network where communities can no longer be detected. This step is reached either because the network is composed of isolated nodes, or because we choose to stop when communities have sizes smaller than a certain predetermined value $S_{stop}$. Hereafter we will stop the iterations when observing communities with sizes smaller than $S_{stop}=3$.

Since at each iteration we remove a node from the existing network and thus affect the detected community structure, we reapply at each step the chosen community detection algorithm. This poses some computational constraints in terms of size of the networks as well as in terms of the community detection algorithm computational efficiency. Moreover, since community detections algorithms usually involve some level of stochasticity, we apply the whole dismantling procedure $M$ times in order to assess what is the robustness of the obtained results. 

In this work we will mostly consider the community detection algorithm introduced in  \cite{fast}. This algorithm belong to the wide class of those algorithms that search communities by maximizing the network modularity. It provides an improvement of the Louvain algorithm \cite{louvain} because it provides (i) a better detection of communities and (ii) faster computational times, which is relevant to our case given the need to rerun the community detection at each iteration of the procedure.

\section*{Authors contribution}
FM and SM designed research; FM and SM analyzed the data and performed numerical simulations; FM and SM analyzed results and wrote the paper. All authors read and approved the final manuscript.
\section*{Data availability}
All the datasets used in our analysis are publicly available and properly referenced in the text, apart from the airport network that comes from a proprietary dataset that cannot be shared without authorization of the Eurocontrol authority.
\section*{Code availability}
The code for replication of the results of the paper will be released upon publication.
\section*{Competing interests}
The authors declare no competing interests. 
\bibliographystyle{IEEEtran}
%

\end{document}